# Localized atomic vibrations caused by point impurity in long chains of noble gas atoms adsorbed in outer grooves of carbon nanobundle

E.V. Manzhelii, S.B. Feodosyev

*B. Verkin Institute for Low Temperature Physics and Engineering of the National Academy of Sciences of Ukraine, 47 Nauky Ave., Kharkiv 61103, Ukraine*

*E-mail:* emanzhelii@ilt.kharkov.ua

**Abstract**

The characteristics of discrete vibrational levels caused by a point three-parameter substitutional impurity in long linear chain of inert gas atoms adsorbed in groove on the surface of carbon nanobundle are studied. The impurity atom differs from the atoms of the chain in the following parameters: the mass, the parameter of interaction with neighboring atoms and the parameter of interaction with the substrate. Analytical expressions for the frequencies of the localized vibrations and the intensities of these vibrations are obtained. The conditions for the existence of localized vibrations both below and above the band of the quasi-continuous spectrum of the adsorbed chain are also obtained.

**Introduction**

Quasi-low-dimensional systems are of interest due to their unusual properties, including its vibrational properties. At present, along with stable two-dimensional atomic structures (graphene), macroscopically long one-dimensional adsorbed atomic structures have also been obtained. Due to the instability of infinite linear chains [1], the problem of creation of macroscopically long adsorbed atomic and molecular chains has not been solved for a long time. Nanotube bundles offer the unique opportunity to solve this problem. The grooves between the nanotubes on the surface of the bundles are the almost ideal matrices for the adsorption of long linear chains of atoms. Chains of adsorbed inert gases are formed in the outer channels of the carbon nanobundle when a small amount of gas is adsorbed on the surface of the nanobundle at low temperatures. Also, chains of atoms can be adsorbed inside nanotubes [2] and in internal channels of nanotube bundles [3]. Direct evidence of the one-dimensional and periodic nature of the structures formed by adsorbed gases in grooves on the surface of a nanobundle provide the neutron diffraction experiments [4, 5]. The linear and periodic nature of these structures is indirectly confirmed by the studies of their heat capacity [6-10]. Also, analysis of the heat capacity and the thermal expansion of xenon atom chains, adsorbed in outer grooves of nanobundle, led to the conclusion that these chains fragment with increasing temperature [10].

Note that the study of the heat capacity of structures formed by helium atoms in the internal channels of nanobundles indirectly confirms their linear and periodic structure [3]. It is natural to be interested in theoretical studies of the vibrational characteristics and the heat capacity of chains of inert gas atoms adsorbed in grooves on the surface of nanobundles [11-16].

The problem of the vibrational characteristics and the heat capacity of an ideal chain of inert gas atoms adsorbed in a groove on the surface of a carbon nanobundle is studied in [11]. Due to the complexity of the crystal field along the groove between nanotubes, the choice of a model to describe the vibrational characteristics of adsorbed chains seems quite difficult. In [11], two models were consistently proposed for description of the ideal chain of inert gas atoms adsorbed in a groove on the surface of a nanobundle. The first model considers a chain of atoms applied to the surface of an fcc crystal (plane (111)). The period of the chain coincides with the period of the crystal field along the chain. It is shown that the spectral density of the applied chain coincides more nearly accurately with the spectral density of the chain in an external periodic field. The period of this field coincides with the period of the chain. In this case, the quasi-continuous spectrum of the chain begins with a certain initial frequency $\omega_{min}$, which is determined by the interaction of the chain atoms with the substrate. The width of this spectrum is determined by the interaction of the atoms of the chain with each other. We also showed in [11] that at frequency $\omega > \omega_{min}$ the structure of the carbon substrate does not significantly affect the nature of the phonon spectrum of the applied chain of noble gas atoms. In particular, this is due to the fact that the spectrum of inert gas atoms is located in the Debye region of the spectrum of the carbon substrate. A significant disadvantage of this model is the low stability of the applied chain of atoms. This fact is evidenced by the high values of the root-mean-square displacements of the transverse vibrations. This disadvantage of the model is eliminated by considering a chain of atoms implanted into the surface layer of an fcc crystal. The heat capacity of a chain of Xe atoms obtained within the framework of this model coincides more nearly accurately with the experimentally obtained heat capacity of a chain of Xe atoms in a groove on the surface of a nanobundle [8, 11]. In the model of a chain of atoms implanted into the surface layer of the fcc crystal, the spectral densities of the chain of atoms shows three-dimensional behavior at low frequencies. This behavior ensures the stability of the chain. The spectral density values in this region are very low. At frequencies $\omega \geq \omega_{min}$, the spectral densities of longitudinal vibrations and vibrations normal to the substrate plane have a character close to one-dimensional one, and transverse vibrations in the substrate plane are quasi-localized.

The models described above enable us to consider a chain of inert gas atoms adsorbed in a groove between nanotubes on the surface of a carbon nanobundle as a chain of atoms in a periodic external crystal field. Therefore, the problem of the vibrational spectra of adsorbed chains without defects is reduced to a problem with two parameters. These parameters are the initial frequency $\omega_{min}$ and the maximum frequency $\omega_{max}$ of the quasi-continuous phonon spectrum of the chain. We consider this approach to be basic for studying the vibrational properties of adsorbed chains containing defects. Changes in the phonon spectrum of atomic chains containing defects are not only of independent interest, but also can affect their low-temperature heat capacity. In [11] it is shown that near the temperature T=0**K** the heat capacity temperature dependence of the adsorbed chain of atoms has the near-exponential shape. With increasing temperature, the low-temperature heat capacity curve takes the near-linear shape. It is shown that the width of the near-exponential heat part of the heat capacity curve increases with increasing the ratio $\omega_{min}/\omega_{max}$. It should be noted that the experimentally obtained curves of the temperature dependence of the low-temperature heat capacity of helium atom chains in the internal channels of nanobundles have this shape [3]. The width of the near-exponential part of heat capacity temperature dependence curve and the slope of its linear part depend on the method of nanobundle preparation [3]. In particular, the nanobundles differed in their impurity composition. In [12-14] it is shown, that when the frequency of vibrations localized on chain defects is below the initial frequency of its spectrum $\omega_{min}$, the near-exponential portion of its low-temperature heat capacity is reduced. These facts explain the interest in the influence of impurities in the adsorbed chain of atoms on its spectrum. The aim of this work is to obtain analytical expressions for the frequencies, conditions of existence and intensities of vibrations localized on a point substitution impurity in a chain of inert gas atoms adsorbed in a groove on the surface of a carbon nanobundle.

**Vibrational characteristics of an adsorbed chain with a defect**

We consider a chain of inert gas atoms with an isolated substitutional impurity adsorbed on a nanobundle. The parameters of the defect are: the mass of the impurity $m_d = m(1+\varepsilon)$, the parameter of force interaction of the impurity with neighboring atoms $\alpha_{di} = \alpha_i(1+\eta_i)$, and the parameter of interaction of the chain atom with the substrate $f_{di} = f_i(1+\xi_i)$. Here $m$ is the mass of the chain atom, $\alpha_i$ is the parameter of the force interaction between the chain atoms, $f_i$ is the parameter of the interaction of the chain atom with the substrate. The index $i = l, \tau$ corresponds to longitudinal and transverse vibrations. As noted in the introduction, a chain of inert gas atoms adsorbed in a groove on the surface of a nanotube bundle can be considered as a chain in an external crystal field [12–14]. In this paper, we restrict ourselves to the interaction of only the nearest neighbors. This approximation is sufficient for a chain of inert gas atoms.

Let us write down the dispersion relation of an ideal chain in an external field, taking into account the only the nearest neighbors interaction:

$$\lambda_i(k) \equiv \omega_i^2(k) = \frac{f_i}{m} + \frac{4\alpha_i}{m}\sin^2\frac{ka}{2}. \tag{1}$$

Here $\lambda \equiv \omega^2$ is the square of the vibration frequency, $a$ is the interatomic distance in the chain, $k$ is the quasi-wave vector. Note that the symmetry condition for the elastic moduli tensor should be applied to the system as a whole (including not only the chain, but also the nanotube bundle), since it is the interaction of the chain with the nanotube atoms that ensures its stability. Therefore, the transverse vibrations of the chain atoms are not flexural vibrations with dispersion $\lambda_\tau(k) \sim k^4$ in the long-wave region.

Due to the symmetry, crystal structures with point defects can be effectively studied using the expansion of the displacements of crystal lattice atoms in spherical waves. Let's briefly look at the mathematical method of Jacoby matrices [17-22] and the vibrational properties of an ideal chain of atoms in an external field.

Let us consider a monatomic chain with a substitution impurity. The chain is in an external field. We will assume that the impurity atom is located at the origin of coordinates. From the symmetry of the system, it is obvious that the space of displacements of the atoms of the chain can be represented as a direct sum of orthogonal cyclic subspaces: $H = H^{(-)} \oplus H^{(0)}$. In each of the subspaces we select a generating vector $\mathbf{h}_0$. The subspace $H^{(0)}$ is the subspace of in-phase displacements of the chain atoms. The subspace $H^{(-)}$ is the subspace of antiphase displacements. For the subspace $H^{(0)}$ the generating vector is $\mathbf{h}_0^{(0)} = |0|1\rangle$. Vector $\mathbf{h}_0^{(0)}$ is the unit displacement of an atom located at the origin of coordinates. The generating vector $\mathbf{h}_0^{(-)} = \frac{1}{\sqrt{2}}\left|\begin{matrix}-a & 1\\ a & -1\end{matrix}\right\rangle$ of the subspace $H^{(-)}$ is the anti-phase unit displacement of the nearest neighbors of the impurity atom. In this subspace, the displacement of the impurity atom is equal to zero. The subspaces $H^{(-)}$ and $H^{(0)}$ are the linear spans stretched over the sequences of vectors $\{\mathbf{L}^n\mathbf{h}_0^{(-)}\}_{n=0}^{\infty}$ and $\{\mathbf{L}^n\mathbf{h}_0^{(0)}\}_{n=0}^{\infty}$, respectively:

$$\{\mathbf{L}^n\mathbf{h}_0\}_{n=0}^{\infty} = \mathbf{h}_0, \mathbf{L}\mathbf{h}_0, \ldots \mathbf{L}^n\mathbf{h}_0\ldots \tag{2}$$

**L** is the dynamic operator acting on space of displacements of lattice atoms, $n$ is the atom number. The dynamic operator **L** has the form:

$$\mathbf{L} = \frac{\Phi_{ik}(r,r')}{\sqrt{m(r)m(r')}}. \tag{3}$$

Here $r, r'$ are the coordinates of the chain atoms, $\Phi_{ik}(r,r')$ is the matrix of force constants. The dynamic operator $\mathbf{L}_{ch}$ for an adsorbed chain without defects has the form:

$$\mathbf{L}_{ch}(r,r') = \frac{1}{m}\begin{pmatrix} f_l + 2\alpha_l & 0 & 0 \\ 0 & f_\tau + 2\alpha_\tau & 0 \\ 0 & 0 & f_\tau + 2\alpha_\tau \end{pmatrix} \cdot \delta_{r,r'} - \frac{1}{m}\begin{pmatrix} \alpha_l & 0 & 0 \\ 0 & \alpha_\tau & 0 \\ 0 & 0 & \alpha_\tau \end{pmatrix} \cdot \left(\delta_{r,r'+a} + \delta_{r,r'-a}\right). \tag{4}$$

As a result of orthonormalization [20] of sequences of vectors $\{\mathbf{L}^n\mathbf{h}_0^{(-)}\}_{n=0}^{\infty}$ and $\{\mathbf{L}^n\mathbf{h}_0^{(0)}\}_{n=0}^{\infty}$ we obtain bases $\{\mathbf{h}_n^{(-)}\}_{n=0}^{\infty}$ and $\{\mathbf{h}_n^{(0)}\}_{n=0}^{\infty}$. In each of these bases, the operator **L** represents the Jacoby matrix. The elements of the Jacoby matrix have the form:

$$\mathbf{L}_{nn'} = a_n \delta_{nn'} + b_n \left(\delta_{n,n'+1} + \delta_{n+1,n'}\right). \tag{5}$$

Here and in what follows, the absence of superscripts indicating the displacement subspace in expressions will mean that these expressions are true in each of the subspaces. The absence of the subscript $i$ in expressions means that the expressions are true for each of the branches of vibrations. The matrix elements of the Jacoby matrix have an important property. As $n$ tends to the infinity, the matrix elements tend to their limiting values:

$$a = \frac{2\alpha + f}{m}, \quad b = \frac{\alpha}{m}. \tag{7}$$

Complete information about the vibrational spectrum of a chain is contained in its Green operator $\mathbf{G}(\lambda) = (\lambda\mathbf{I} - \mathbf{L})^{-1}$ (**I** is the unit operator). In the Jacobi matrix formalism, all matrix elements of the operator $\mathbf{G}(\lambda)$ are expressed through the elements:

$$G_{00}(\lambda) = \left(\mathbf{h}_0, \mathbf{G}(\lambda)\mathbf{h}_0\right). \tag{8}$$

The Green operator element $G_{00}(\lambda)$ can be represented as an infinite fraction

$$G_{00}(\lambda) = \cfrac{1}{a_0 - \lambda - \cfrac{b_0^2}{a_1 - \lambda - \cfrac{b_1^2}{a_2 - \lambda - \cfrac{b_2^2}{\cdots \cfrac{}{a_n - \lambda - \cfrac{b_n^2}{a - \lambda - K_\infty(\lambda)}}}}}} \quad . \tag{9}$$

Here $K_\infty(\lambda)$ is the fraction corresponding to the Jacoby matrix, all elements of which are equal to their limit values. For a chain in an external field, the fraction $K_\infty(\lambda)$ has the form:

$$K_\infty(\lambda) = \frac{\lambda - a + Z(\lambda)\sqrt{(\lambda-a)^2 - 4b^2}}{2b^2} . \tag{10}$$

Here $Z(\lambda) = \Theta(\lambda_{min} - \lambda) + i\Theta(\lambda - \lambda_{min})\Theta(\lambda_{max} - \lambda) - \Theta(\lambda - \lambda_{max})$, $\Theta(\lambda)$ is the Heaviside theta function. The values $\lambda_{min}$ and $\lambda_{max}$ are squares of the minimum and maximum frequencies of the quasi-continuous spectrum of vibrations, respectively ($\lambda_{min} = a - 2b$, $\lambda_{max} = a + 2b$). The spectral density $\rho(\lambda) = \pi^{-1} \operatorname{Im} G_{00}(\lambda)$ can be obtained in each of the subspaces. The density of the vibrational states $g(\lambda)$ is equal to the arithmetic mean of the spectral densities in the cyclic subspaces.

Consider an adsorbed chain of atoms with a local defect. Dynamic operator of a perturbed system is $\tilde{\mathbf{L}} = \mathbf{L} + \mathbf{\Lambda}$, here is $\mathbf{\Lambda}$ the perturbation operator. In the future, the "wave" over the values will indicate the presence of a defect in the chain. In each of the subspaces, the squares of the frequencies of vibrations localized on the defect $\lambda_d$ are the roots of the denominator of the function $\tilde{G}_{00}(\lambda)$ of the chain with a defect. However, not all roots of the denominator of the function $\tilde{G}_{00}(\lambda)$ are squared frequencies of localized vibrations. The intensities corresponding to these roots must be positive. The residues at these poles are the intensities of these vibrations:

$$\mu_{d0} \equiv \operatorname*{res}_{\lambda = \lambda_d} \tilde{G}_{00}(\lambda) . \tag{11}$$

Also, the squared frequency of the vibration localized on the defect is the root of the Lifshitz equation:

$$\operatorname{Re} G_{00}(\lambda) = 1/\Lambda_{00} . \tag{12}$$

We emphasize that $G_{00}(\lambda)$ is the element of the Green operator of a chain without defects and $\Lambda_{00}$ is:

$$\Lambda_{00} = (\mathbf{h}_0, \mathbf{\Lambda}\mathbf{h}_0). \tag{13}$$

Thus, from the form of the Green operator element of a chain without defects $G_{00}(\lambda)$, one can conclude whether vibrations localized on defects will be formed with or without a threshold.

## DISCRETE VIBRATIONS CAUSED BY THREE-PARAMETER SUBSTITUTIONAL IMPURITY

### a. Localized vibrations in the subspace of antiphase displacements.

Let us consider the vibrations localized on an impurity atom in the subspace of antiphase displacements $H^{(-)}$. In what follows, the superscript $(-)$ will denote quantities in the subspace $H^{(-)}$. In this subspace, the only element of the Jacobi matrix that depends on the defect parameter is:

$$\tilde{a}_0^{(-)} = \frac{\alpha + \alpha_d + f}{m}. \tag{14}$$

All other elements of the Jacoby matrix are equal to their limiting values. Therefore, in this subspace, the only non-zero element of the perturbation operator $\Lambda$ is:

$$\Lambda_{00}^{(-)} = \frac{\eta \alpha}{m}. \tag{15}$$

The Green's operator element $\tilde{G}_{00}(\lambda)$ in this subspace can be written as:

$$G_{00}^{(-)} = -\frac{\lambda - \left(a + 2\Lambda_{00}^{(-)}\right) + Z(\lambda)\sqrt{\left|(\lambda - a)^2 - 4b^2\right|}}{2\Lambda_{00}^{(-)}\left(\lambda - \lambda_d^{(-)}\right)}, \tag{16}$$

From (16) we obtain the squared frequency of localized vibrations:

$$\lambda_d^{(-)} = \frac{b^2 + \left(\Lambda_{00}^{(-)}\right)^2}{\Lambda_{00}^{(-)}} + a. \tag{17}$$

In the subspace $H^{(-)}$ the following expression for the intensity of vibrations localized on the impurity $\mu_{d0}^{(-)}$ is obtained:

$$\mu_{d0}^{(-)} = \underset{\lambda \to \lambda_d^{(-)}}{res}\, G_{00}^{(-)}\left(\lambda, \Lambda_{00}^{(-)}\right) = -\frac{b^2 - \left(\Lambda_{00}^{(-)}\right)^2}{2\left(\Lambda_{00}^{(-)}\right)^2} - \frac{Z(\lambda)}{2\Lambda_{00}^{(-)}}\left|\frac{b^2 - \left(\Lambda_{00}^{(-)}\right)^2}{\Lambda_{00}^{(-)}}\right|. \tag{18}$$

From the condition $\mu_{d0}^{(-)} > 0$ we obtain the values of the thresholds for the formation of localized vibrations. At $\lambda_d^{(-)} > \lambda_{max}$ localized vibrations exist only at positive values of $\Lambda_{00}^{(-)}$. The condition for the existence of localized vibrations is $\Lambda_{00}^{(-)} > |b|$. In the region of squared frequencies $\lambda_d^{(-)} < \lambda_{min}$, localized vibrations are formed only at negative values of $\Lambda_{00}^{(-)}$. The attenuation of the amplitude of localized vibrations with increasing n is characterized by the quantity $\mu_n(\lambda, \Lambda) = \underset{\lambda \to \lambda_d}{res}\, G_{nn}(\lambda, \Lambda) = \mu_{d0} P_n^2(\lambda_d)$ [23]. The functions $P_n(\lambda)$ are polynomials of degree $n$ satisfying the recurrence relation:

$$b_n P_{n+1}(\lambda) = (\lambda - a_n) P_n(\lambda) - b_{n-1} P_{n-1}(\lambda). \tag{19}$$

The initial conditions for $P_n(\lambda)$ are:

$$P_{-1}(\lambda) = 0, \quad P_0(\lambda) = 1. \tag{20}$$

It has been shown that polynomials $P_n^2(\lambda_d)$ form a decreasing geometric progression [20–23]. It was also shown in [23] that for any local vibrational level the following condition must be satisfied:

$$\sum_{n=0}^{\infty} \mu_n(\lambda_d) = \mu_0(\lambda_d) \sum_{n=0}^{\infty} P_n^2(\lambda_d) = 1. \tag{21}$$

The fulfillment of this condition means that each local vibration corresponds to the splitting of one phonon from the quasi-continuous spectrum. It is easy to verify that the relation (21) is satisfied for the subspace $H^{(-)}$.

The presence of thresholds in the formation of localized states can be illustrated using the Lifshitz equation (12). The element of the Green's operator $G_{00}^{(-)}$ of the chain, without impurities in this subspace, has the form:

$$G_{00}^{(-)}(\lambda) = \frac{2}{\lambda - a - Z(\lambda)\sqrt{|(\lambda_{min} - \lambda)(\lambda - \lambda_{max})|}}. \tag{22}$$

Below in Fig. 1 we present the graphical solution of the Lifshitz equation (12) in the subspace of antiphase displacements of the atoms of the chain.

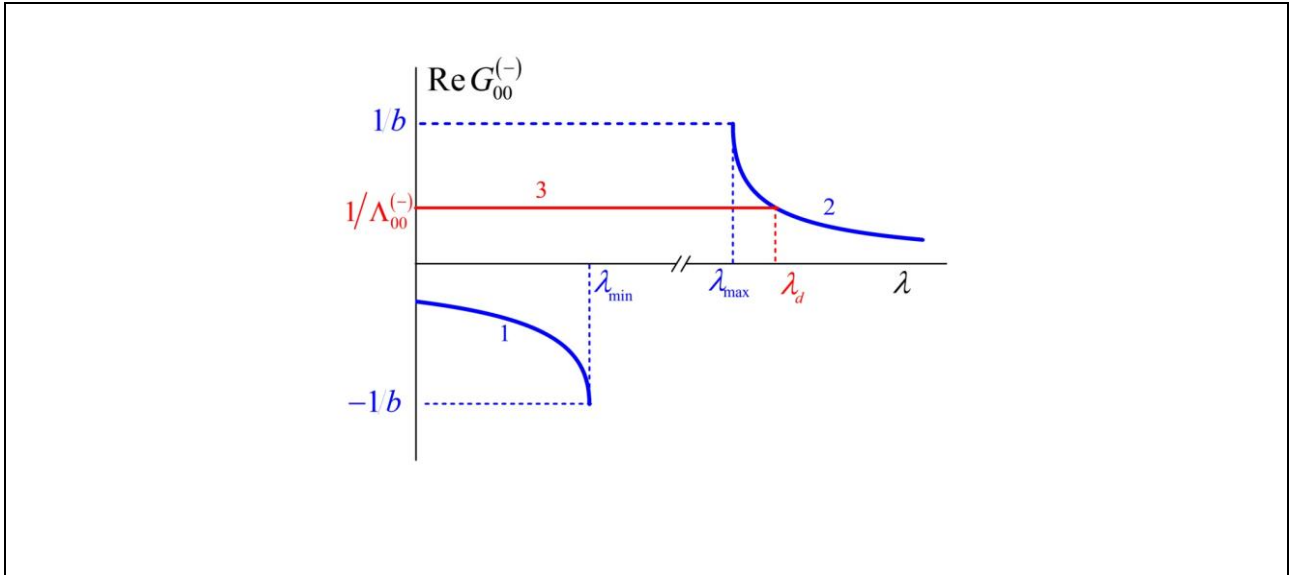

Fig.1: The graphical solution of the Lifshitz equation in the subspace $H^{(-)}$. The curves 1, 2 are the real parts of the $G_{00}^{(-)}$ element of the Green's operator. The straight line 3 is the reciprocal of the $\Lambda_{00}^{(-)}$ element of the perturbation operator.

Thus, due to the fact that in the subspace of antiphase displacements the impurity atom is motionless, discrete vibrations are localized on its nearest neighbors. Localized vibrations in this subspace arise starting from the values of the change in the interaction parameter between the atoms of the chain and the impurity $|\eta|>1$. Each local vibration corresponds to the splitting of one phonon from the quasi-continuous spectrum.

**b. Localized vibrations in the subspace of in-phase displacements.**

Let us consider localized vibrations caused by a point impurity in the subspace of in-phase displacements $H^{(0)}$. As was said above, the subspace $H^{(0)}$ is generated by the unit displacement vector of the impurity atom $\mathbf{h}_0^{(0)} = |0|1\rangle$. In what follows, the superscript (0) will denote quantities in the subspace $H^{(0)}$. In this subspace, only the elements of the Jacobi matrix $\tilde{a}_0^{(0)}$, $\tilde{b}_0^{(0)}$ and $\tilde{a}_1^{(0)}$ depend on the defect parameters. In the problem we are considering, these elements have the form:

$$\tilde{a}_0^{(0)} = \frac{2\alpha(1+\eta)}{m(1+\varepsilon)} + \frac{f(1+\xi)}{m(1+\varepsilon)} = a + 2\frac{\alpha}{m}\left(\frac{\eta-\varepsilon}{1+\varepsilon}\right) + \frac{f}{m}\left(\frac{\xi-\varepsilon}{1+\varepsilon}\right), \tag{23}$$

$$\tilde{b}_0^{(0)} = \sqrt{2}b\frac{1+\eta}{\sqrt{1+\varepsilon}}, \tag{24}$$

$$\tilde{a}_1^{(0)} = a_1 + \frac{\alpha\eta}{m}. \tag{25}$$

For an ideal chain $b_0^{(0)} = \sqrt{2}b$, all other elements of its Jacoby matrix are equal to their limit values. Hence:

$$\Lambda_{00}^{(0)} = 2\frac{\alpha}{m}\left(\frac{\eta-\varepsilon}{1+\varepsilon}\right) + \frac{f}{m}\left(\frac{\xi-\varepsilon}{1+\varepsilon}\right) \tag{26}$$

The element of the Green operator $\tilde{G}_{(00)}^{(0)}$ in this subspace has the form:

$$\tilde{G}_{(00)}^{(0)}(\lambda) = \frac{1}{\lambda - \tilde{a}_0^{(0)} - \dfrac{\left(\tilde{b}_0^{(0)}\right)^2}{\lambda - \tilde{a}_1^{(0)} - b^2 K_\infty(\lambda)}}. \tag{27}$$

As a result we obtain:

$$\tilde{G}_{00}^{(0)}(\lambda) = \frac{g_1 g_2}{R(\lambda)}, \tag{28}$$

where $g_1 = \lambda - 2\tilde{a}_1^{(0)} + a - Z(\lambda)\sqrt{(\lambda-a)^2 - 4b^2}$,

$g_2 = \left(\lambda - \tilde{a}_0^{(0)}\right)\left(\lambda - 2\tilde{a}_1^{(0)} + a\right) - 2\left(\tilde{b}_0^{(0)}\right)^2 + Z(\lambda)\left(\lambda - \tilde{a}_0^{(0)}\right)\sqrt{(\lambda-a)^2 - 4b^2}$,

$R_i(\lambda) = \left[\left(\lambda - \tilde{a}_0^{(0)}\right)\left(\lambda - 2\tilde{a}_1^{(0)} + a\right) - 2\left(\tilde{b}_0^{(0)}\right)^2\right]^2 - \left(\lambda - \tilde{a}_0^{(0)}\right)^2\left[(\lambda-a)^2 - 4b^2\right]$.

We see that $R(\lambda)$ is a cubic polynomial. To overcome the difficulties caused by this fact, the "two-momentum" approximation was proposed and substantiated in [23-24]. This approximation is based on the fact that Green's functions converge extremely quickly at the frequencies located outside the quasi-continuous spectrum band. In this approximation, we assume that all elements of the Jacoby matrix except the elements $\tilde{a}_0^{(0)}$ and $\tilde{b}_0^{(0)}$ are equal to their limit values. In this approximation:

$$\tilde{G}_{00}^{(0)}(\lambda) = \frac{q_1(\lambda) q_2(\lambda)}{Q(\lambda)}, \tag{29}$$

where $q_1(\lambda) = \lambda - a - Z(\lambda)\sqrt{(\lambda-a)^2 - 4b^2}$,

$q_2(\lambda) = \left(\lambda - \tilde{a}_0^{(0)}\right)(\lambda - a) - 2\left(\tilde{b}_0^{(0)}\right)^2 + Z(\lambda)\left(\lambda - \tilde{a}_0^{(0)}\right)\sqrt{(\lambda-a)^2 - 4b^2}$

$Q(\lambda) = 4\left(b^2 - \left(\tilde{b}_0^{(0)}\right)^2\right)\lambda^2 + 4\left(a\left(\tilde{b}_0^{(0)}\right)^2 + \tilde{a}_0^{(0)}\left(\left(\tilde{b}_0^{(0)}\right)^2 - 2b^2\right)\right)\lambda + 4\left(\left(\tilde{a}_0^{(0)}\right)^2 b^2 - a\tilde{a}_0^{(0)}\left(\tilde{b}_0^{(0)}\right)^2 + \left(\tilde{b}_0^{(0)}\right)^4\right)$

Thus, the squares of the frequencies of localized vibrations are the roots of the quadratic trinomial and have the form:

$$\lambda_{1,2} = \frac{-B \pm \sqrt{B^2 - 4AC}}{2A}, \tag{30}$$

here $A = b^2 - \left(\tilde{b}_0^{(0)}\right)^2$, $B = a\left(\tilde{b}_0^{(0)}\right)^2 + \tilde{a}_0^{(0)}\left(\left(\tilde{b}_0^{(0)}\right)^2 - 2b^2\right)$, $C = \left(\tilde{a}_0^{(0)}\right)^2 b^2 - a\tilde{a}_0^{(0)}\left(\tilde{b}_0^{(0)}\right)^2 + \left(\tilde{b}_0^{(0)}\right)^4$.

Accordingly, the condition for the existence of localized oscillations has the form:

$$\mu_{d0}^{(0)}\left(\lambda_{1,2}\right) = \frac{q_1\left(\lambda_{1,2}\right) q_2\left(\lambda_{1,2}\right)}{4\left(2A\lambda_{1,2} + B\right)} > 0. \tag{31}$$

The expression (31) is quite cumbersome. In this case, it is much easier to find the conditions for the existence of quasi-local vibrations from the Lifshitz equation (12). The element $G_{00}^{(0)}$ of the Green's operator of a chain without defects in this subspace has the form:

$$G_{00}^{(0)} = -\frac{1}{Z(\lambda)\sqrt{\left|(\lambda_{min} - \lambda)(\lambda_{max} - \lambda)\right|}} \tag{32}$$

It is evident that for any arbitrarily small value of the element $\Lambda_{00}^{(0)}$ of the perturbation operator, the Lifshitz equation (12) has a solution. We present the graphical solution of the Lifshitz equation (12) (Fig. 2):

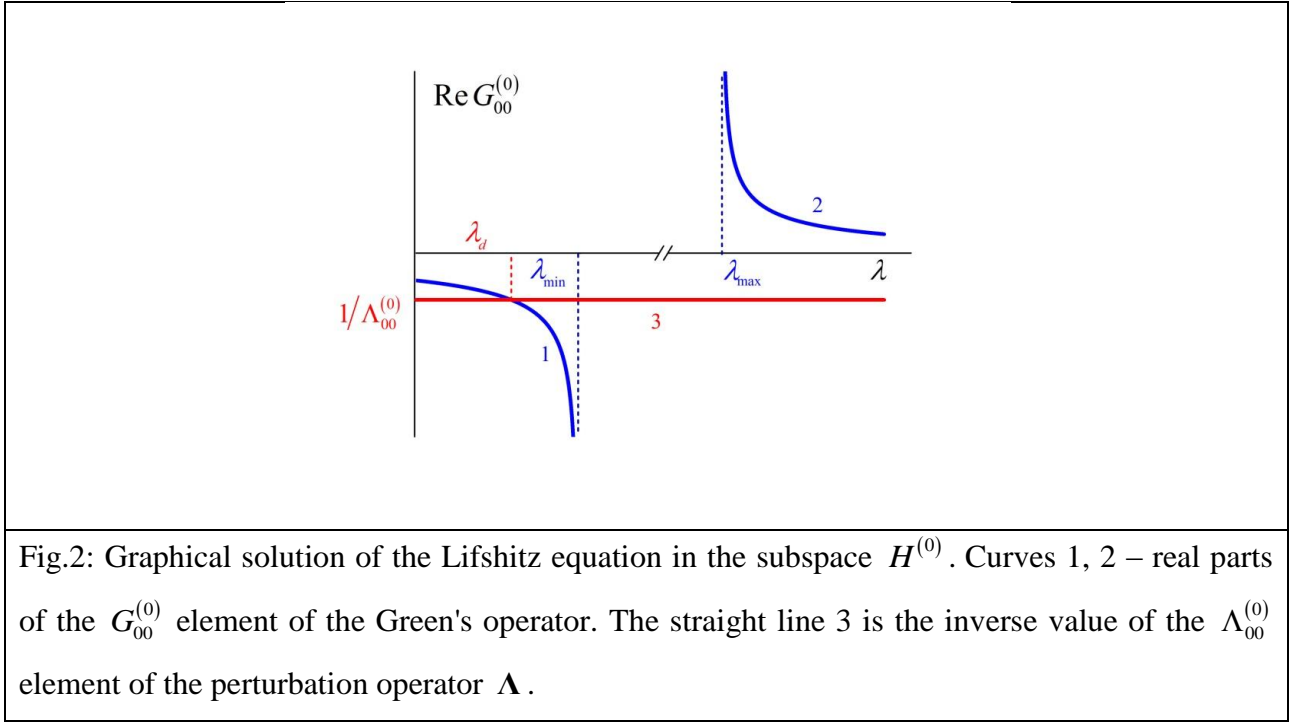

Fig.2: Graphical solution of the Lifshitz equation in the subspace $H^{(0)}$. Curves 1, 2 – real parts of the $G_{00}^{(0)}$ element of the Green's operator. The straight line 3 is the inverse value of the $\Lambda_{00}^{(0)}$ element of the perturbation operator $\Lambda$.

As a result, we see that in the subspace of in-phase displacements, the discrete vibrations are localized on the impurity atom. These vibrations are formed without a threshold.

**Conclusions**

Thus, in this work it is shown that an isolated three-parameter substitutional point impurity in a chain of inert gas atoms adsorbed in a groove between two nanotubes on the surface of a nanobundle leads to the emergence of localized vibrations with frequencies both below and above the quasi-continuous spectrum band. These vibrations can be localized both on the impurity itself and on the nearest neighbors of the impurity atom. It is shown that oscillations localized on impurities are formed without a threshold, whereas for the occurrence of oscillations localized on the nearest neighbors of the impurity atom, it is necessary to overcome a threshold. The analytical expressions for the frequencies of localized vibrations, as well as, for the threshold for the occurrence of localized vibrations, and for their intensities are obtained. The analytical expressions for the characteristics of localized vibrations expand the possibilities of experimental study of the parameters of interaction of substrate atoms with chain atoms and with impurity atoms. Also, the obtained analytical expressions give the possibility to analyze the influence of impurities on the shape of the low-temperature heat capacity curve of atomic chains in the outer grooves of nanobundle. It is significant that in the case of a linear chain, the isolated impurity approximation is applicable at impurity concentrations that are significantly higher than in the case of three-dimensional structures [14, 25, 26].

**Acknowledgments**

The authors are grateful to E.S. Syrkin, A.A. Zvyagin and A.S. Kovalev for the fruitful discussion.